\documentclass[aps,prl,twocolumn,showpacs]{revtex4}
\usepackage{epsfig}
\usepackage{bm}
\def\be{\begin{equation}}
\def\ee{\end{equation}}

\begin{document}
\title{Entanglement Energetics at Zero Temperature}
\author{Andrew N. Jordan and Markus B\"uttiker}
\affiliation{D\'epartement de Physique Th\'eorique, Universit\'e de Gen\`eve,
        CH-1211 Gen\`eve 4, Switzerland}
\date{14 November, 2003}

\begin{abstract}
We show how many-body ground state entanglement information may 
be extracted from sub-system energy measurements at zero temperature. 
Generically, the larger the measured energy fluctuations are, 
the larger the entanglement is.
Examples are given with the two-state system and the harmonic
oscillator. Comparisons made with recent qubit experiments
show this type of measurement provides another method to quantify
 entanglement with the environment. 
\end{abstract}

\maketitle

A many-body quantum system is cooled to zero temperature
so that it is forced into its overall nondegenerate ground state.
We discuss the measurement of a sub-system Hamiltonian 
and demonstrate that it can be found 
in an excited state with a probability that depends 
on the coupling to its environment.  This non-intuitive 
result is a pure quantum phenomenon: it is 
a consequence of entanglement \cite{sch} of the sub-system 
with the environment. In fact, we demonstrate that knowledge of 
the probability to find the system in an excited state can
be used to determine the degree of entanglement of the sub-system 
and bath. Consequently, simple systems with well known 
isolated quantum mechanical
properties (such as the two-state system and harmonic oscillator) 
become ``entanglement-meters''. 

There is growing interest of
ground-state entanglement in condensed matter physics.  
Theoretical works on ground state
entanglement have addressed entropy scaling in harmonic
networks \cite{ms}, 
spin-spin entanglement in quantum spin chains \cite{arnesen}
and quantum phase transitions \cite{fazio,nielsen2}.  
Entanglement properties of the ground state are also 
essential in the field of adiabatic quantum computing \cite{adiabatic}.
Furthermore it is interesting to link other ground state properties 
like the persistent current of small mesoscopic rings \cite{pc,pc2},
or of doubly connected Cooper pair boxes \cite{cpbox1,cpbox2,qubit},
or the occupation of resonant states \cite{resonant}
to the zero-temperature entanglement energetics. 


We consider a general Hamiltonian
$H= H_s + H_{c}+ H_{E}$, that couples $(c)$ the system $(s)$ we are
interested in to a quantum environment $(E)$ such as a network
of harmonic oscillators \cite{rmp}.
The lowest energy separable state is 
$\vert \Omega\rangle= \vert 0\rangle_s\vert 0\rangle_{E}$, 
where $\vert 0\rangle_{\{s,E\}}$ are the  
lowest uncoupled energy state of both systems.
However, if the system Hamiltonian and the total Hamiltonian do not
commute (which is the generic situation), then $\vert \Omega\rangle$
is not an energy eigenstate of the total Hamiltonian.  Thus, 
there must be a lower energy eigenstate ($\vert 0
\rangle$) of the total Hamiltonian which is by definition an entangled state.
Because time evolution is governed by the full
Hamiltonian, the ground state expectation of any operator with 
no explicit time dependence will have no time evolution,
insuring that any measurement is static in time.
This situation is in contrast to the usual starting point of assuming that
the initial state is a separable state
and studying how it becomes entangled.
The reduced density operator of the system is given
by tracing out the environmental degrees of freedom, 
$\rho = {\rm Tr}_E \vert 0 \rangle\langle 0 \vert$.
Assuming the full state of the whole
system is pure, the reduced density matrix contains all accessible
system information, including
entanglement of the system with its environment.
Because repeated measurements of $H_s$ will
give different energies as the sub-system is not in an energy
eigenstate, we are interested in a complete description of 
the statistical energy fluctuations.  These fluctuations
may be described in two equivalent ways.
The first way is to find the diagonal density matrix
elements in the basis where $H_s$ is diagonal.  These elements 
represent the probability to measure a particular excited state of $H_s$.
A second way is to find all energy cumulants. 
A cumulant of arbitrary order may be calculated from the sub-system
energy generating function, 
$Z(\chi)=\langle \exp (-\chi H_s)\rangle$ 
(as always, $\langle {\cal O} \rangle= {\rm Tr} \rho {\cal O}$)
so that the n$^{th}$ energy cumulant is given by
\be
\langle\langle H_s^n\rangle \rangle
 = (-)^n \frac{d^n}{d \chi^n} \ln Z(\chi)\Big\vert_{\chi=0}\; .
\label{cumdef}
\ee
These cumulants give information about the measured energy distribution around
the average.  

Before proceeding to calculate these energy fluctuations, 
we ask a general question about entanglement.
Given the energy distribution function (the diagonal
matrix elements of the density matrix only), can anything be said in
general about the purity or entropy of the state?
Surprisingly, because we are given the additional information that
we are at zero temperature, the answer is yes.
If we ever measure the sub-system's energy and find
an excited energy, then we know the state is entangled.
Though this statement alone links energy fluctuations with
entanglement, a further quantitative statement may be made in 
the weak coupling
limit.  The reason for this is the following:  
the assumptions exponentially
suppress higher states, so to first order in the coupling
constant, we can consider a two-state system where
the density matrix has the form 
$\rho_{--}=1-\alpha p$, $\rho_{++}=\alpha p$, 
$\rho_{+-}=\rho_{-+}^\ast = \alpha c$. 
For vanishing coupling constant $\alpha=0$, this just
gives the density matrix for the separable state. 
The linear dependence of $\rho$ on $\alpha$ holds
to first order for the model systems considered below
and is the entanglement contribution.
If one measures the diagonal elements of
$\rho$, one obtains $\rho_{--}$ and  
$\rho_{++}$ as the probability to be measured in 
the ground or excited
state (because $\alpha$ is small, there is only a small probability
to find the sub-system in the upper state).
If we now diagonalize $\rho$, the eigenvalues
are $\lambda_\pm = \{ 1- p \, \alpha, p\,  \alpha \} +{\cal O}(\alpha^2)$.  To first order in $\alpha$,
the eigenvalues are the diagonal matrix elements, so
we may (to a good approximation) write the purity or entropy in terms
of these probabilities even if the energy
difference remains unknown.

{\it The Qubit.}
Let us now first evaluate the energy fluctuations 
of a qubit, a two-state system. 
The most general (trace 1) spin density matrix is
$\rho = ({\openone} + \langle
\sigma_x \rangle \sigma_x+ \langle\sigma_y\rangle\sigma_y
+ \langle\sigma_z \rangle \sigma_z)/2$. 
A simple measure of the entanglement is given by the purity,
$ {\rm Tr} \rho^2 = (1/2) (1+X^2+Y^2+Z^2)$, where 
$X_i= \langle \sigma_i \rangle$.
It is well known that $(X,Y,Z)$ form coordinates in the Block sphere.
Purity lies at the surface where $X^2+Y^2+Z^2=1$, whereas 
corruption lies deep in the middle.

We take the system Hamiltonian
\cite{note1} to be $H_s= (\epsilon/2)\, \sigma_z +
(\Delta/2)\, \sigma_x$.
Introducing the frequency $\Omega=\sqrt{\epsilon^2+\Delta^2}/\hbar$
and using the identity
$e^{-i \frac{\beta}{2} {\hat n} \cdot {\hat \sigma}}
= {\rm} I \cos\frac{\beta}{2} - 
i {\hat n} \cdot {\hat \sigma} \sin\frac{\beta}{2}$
with $\beta = \chi \Omega$,
it is straightforward to show
\be
Z(i\chi)  = \cos(\Omega \chi/2) -i 
\frac{\sin (\Omega
\chi/2)}{\Omega} (\epsilon \langle\sigma_z\rangle +
\Delta \langle\sigma_x\rangle) \, .
\label{zs}
\ee
The energy probability distribution may be easily found by Fourier
transforming Eq.~(\ref{zs}), or by tracing in the diagonal basis of the system
Hamiltonian.  The answer may be expressed with only the average
energy,
$\langle H_s\rangle =  \frac{\epsilon}{2} \langle \sigma_z\rangle +
\frac{\Delta}{2} \langle \sigma_x \rangle$
as a sum of delta functions at the system energies $\pm \hbar \Omega/2$ with
weights of the diagonal density matrix elements,
\be
\rho_{++,--} =(1/2) \left[1 \pm \frac{\langle
    H_s\rangle}{\hbar \Omega/2}\right]\, .
\label{probs}
\ee
Clearly, if the spin is isolated from the environment, 
$\langle H_s\rangle = -\hbar\Omega/2$, the ground state energy, the
probability weight to be in an excited state vanishes.
This distribution may also be found from knowledge
of the isolated eigenenergies,
the fact that 
$\langle H_s \rangle = \sum_j E_j \rho_{jj} $, and that Tr$\rho=1$.
This later argument may be 
extended to $n$-state systems given the first $n-1$ moments
of the Hamiltonian and the $n$ eigenenergies.  

\begin{figure}[t]
\begin{center}
\psfig{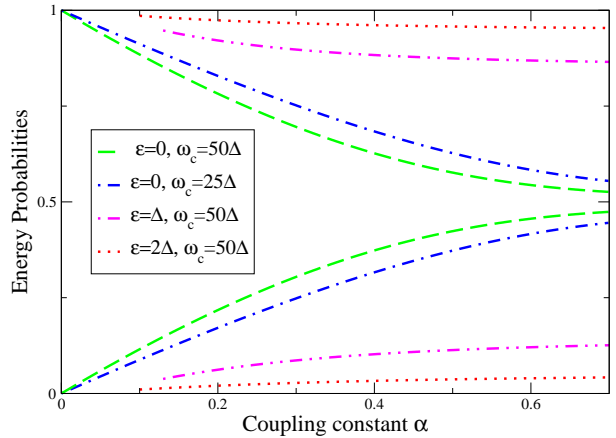}
\caption{Energy probabilities $p_{up}$ and $p_{down}$ 
for the spin-boson problem. With increasing 
coupling to the environment it 
it is more likely to measure the qubit in the excited state.
For the symmetric case $(\epsilon=0)$, we use the Bethe ansatz
solution, while for the general case, we use a perturbative
solution which is only valid for large $\epsilon$ or large $\alpha$.}
\label{qubitprobs}
\end{center}
\end{figure}
{\it Connection with Real Qubits.}
The probability weights depend on the energy parameters $\epsilon$ and
$\Delta$, and the expectation values of the Pauli matrices.
For real qubits produced in the lab, 
these will depend on the environment \cite{spin-boson}.  
Often, we can link the basic phenomena we have been describing 
to physical measurements other than energy.
For example, in a mesoscopic ring
threaded by an Aharonov-Bohm flux $\Phi$, or for the Cooper pair box,
the tunneling matrix element depends on flux $\Delta (\Phi)$. 
The operator one is interested in measuring is given by the normalized 
persistent current \cite{pc}, or the
expectation of $\sigma_x$ (for the symmetric case of $\epsilon=0$),
\be
\langle \sigma_x \rangle  = \frac{I(\Phi)}{I_{0}(\Phi)}\; ,
\label{p1}
\ee
where $I_{0}(\Phi)$ is the uncoupled value of the persistent current.
A common model for environmental effects is given by coupling
the two-state system to a series of harmonic oscillators, 
the spin-boson model \cite{rmp,pc,weiss,spin-boson}.
In Fig.(1) we have plotted the upper and lower occupation
probabilities for the spin-boson model 
as a function of the coupling constant $\alpha$.
For the symmetric case ($\epsilon=0$), we have used the Bethe ansatz
solution \cite{pc,bethe}, while for  $\epsilon$ finite, we have used
the perturbative solution in $\Delta/\omega_c$
which is valid only for larger $\alpha$ or $\epsilon$ \cite{pc}. 
Thus the plot is cut off at a small $\alpha$.
A computational approach calculating the expectation values of the Pauli
matrices over the whole parameter range was given in Ref. \cite{spin-boson}.
For $\epsilon=0$, Eq.~(\ref{p1}) determines also the expectation
value of $H_s$, and thus the weights of the energy distribution
function, Eq.~(\ref{probs}), are directly related
to the persistent current.

Experiments are always carried out at finite temperature, 
and it is important to demonstrate that there exists a 
cross-over temperature to the quantum behavior discussed here.
In the low temperature limit, the thermal occupation probability is 
$p_{up} = e^{-(E_2-E_1)/k T}$.
In the weak coupling limit for the symmetric spin boson
problem, the probability to measure the excited state
scales as $p_{up}= -\alpha \log (\Delta/\omega_c)$ \cite{note2}. 
Setting these factors equal and solving for $T^{\ast}$ yields
\be
k T^{\ast} = -\frac{E_2-E_1}{\log (\alpha \log\frac{\omega_c}{\Delta})}\, .
\label{tcross}
\ee 
Since $T^{\ast}$ scales as the inverse logarithm of the
coupling constant, it is experimentally possible to reach 
a regime where thermal excitation is negligible.

As an order of magnitude estimate, we compare with
the Cooper pair box \cite{cpbox1,cpbox2} which is 
among the most environmentally isolated solid state qubits \cite{qubit}. 
From \cite{cpbox2} which found a $Q \sim 10^{4}$, 
we estimate the quantum probability for the box
to be measured in the excited state as
$p_{up} \sim 10^{-3}-10^{-4}$, which is of same order or larger
than the thermal probability, $p_{th} \sim 10^{-4}$.
Experimentally, $p_{up}$ and $p_{th}$ may be confused by
fitting data with an effective temperature,
$\rho_{th}=\exp(-\beta_{\rm eff} H_s)$ \cite{note3}.
However, one may distinguish true thermal behavior
from the effect described here because $p_{up}$ and
$p_{th}$ depend differently on tunable system parameters
such as $\Delta$.
In fact, $\beta_{\rm eff}$ is an entanglement measure.
This chain of reasoning may be inverted to provide
an estimate for $Q$ given only $p_{up}$.
\begin{figure}[b]
\begin{center}
\leavevmode
\psfig{file=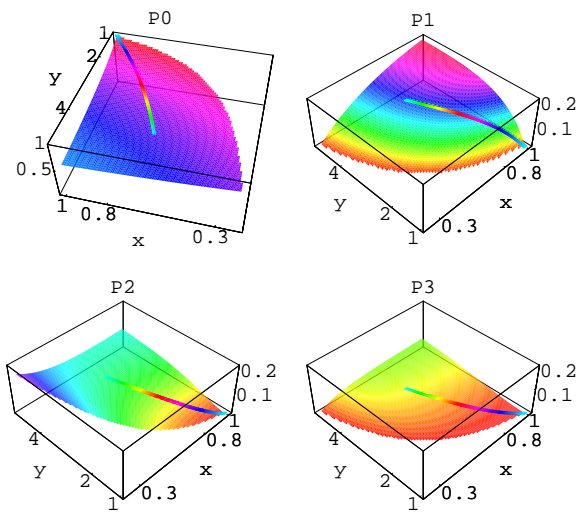,width=8cm}
\caption{The probability to measure a harmonic
oscillator in the ground and first three excited states
as a function of $x$ and $y$ (see text).
The line traces out the behavior of the ohmic bath
as a function of the coupling in the under-damped range.}
\label{bathexcite}
\end{center}
\end{figure}

{\it The Harmonic Oscillator.}
We now consider the entanglement energetics of a harmonic oscillator,
$H_s = p^2/(2 m) + (1/2) m \omega^2 q^2$. 
Since there are an infinite number of states, the problem
is harder.  To simplify our task, we assume a linear coupling
with a harmonic oscillator bath.
This implies that the density matrix is Gaussian so that
environmental information is contained in the second 
moments $\langle q^2 \rangle $ and $\langle p^2 \rangle$ \cite{weiss,ms}, 
\be
\langle q \vert {\rho}\vert q' \rangle = \frac{1}{\sqrt{2 \pi
    \langle q^2 \rangle}} \exp\left\{-\frac{(\frac{q+q'}{2})^2}
{2\langle q^2 \rangle} -  \frac{\langle p^2 \rangle
  (q-q')^2}{2\hbar^2}\right\}.
\label{dm}
\ee
Expectation values of higher powers of $H_s$ are non-trivial because $q$ and
$p$ do not commute.
The purity of the density matrix Eq.~(\ref{dm}) is
\be
{\rm Tr} \rho^2 = \int dq dq' \langle q \vert \rho \vert q'\rangle
\langle q' \vert \rho \vert q\rangle = \frac{\hbar/2}{\sqrt{\langle
    q^2\rangle \langle  p^2\rangle}} \, .
\label{pure}
\ee
The uncertainty relation, 
$\sqrt{\langle q^2\rangle \langle  p^2\rangle} \ge \hbar/2$, 
guarantees that ${\rm Tr} \rho^2 \le 1$, with the
inequality becoming sharp if the oscillator is isolated from the
environment.  As the environment causes greater deviation from the Planck
scale limit, the state loses purity.

The generating function $Z$ may be calculated conveniently by tracing 
in the position basis and inserting a complete set of position states
between the operators,
\be
 Z(\chi) =\int dq dq' \langle q \vert \rho \vert q' \rangle 
\langle q' \vert e^{-\chi H_s} \vert q \rangle .
\label{play}
\ee
The first object in Eq.~(\ref{play}) is the density matrix in position 
representation, given by Eq.~(\ref{dm}).  The second object 
may be interpreted as the uncoupled position-space propagator 
of the harmonic
oscillator from position $q$ to $q'$ in time $-i \hbar \chi$.
%
We find
\be
Z = \left \{ 2 E \,\frac{\sinh \varepsilon\chi}{\varepsilon}
+2 A \,(\cosh \varepsilon \chi -1) + \frac{1+ \cosh \varepsilon \chi}{2} 
\right\}^{-\frac{1}{2}}
\label{zans}
\ee
where
$\varepsilon = \hbar \omega$,
$2 E = m \omega^2 \langle q^2\rangle + \langle p^2\rangle /m$ and
$A =  \langle q^2\rangle \langle p^2\rangle /\hbar^2$.
$E$ is the average energy of the oscillator, while
$A\ge 1$ is a measure of satisfaction of the uncertainty principle.
Eq.~(\ref{zans}) has a pleasing limit for the free particle
$\omega\rightarrow 0$,
\be
Z(\chi)_{free} = \left \{ 1+ \chi \langle p^2 \rangle/m
\right\}^{-\frac{1}{2}}\, ,
\label{zans2}
\ee
which is just the generating function for Wick contractions,
$\langle p^{2 n}\rangle = (2n -1)!!\, (\langle p^2 \rangle)^n$.
Thus, in Eq.~(\ref{zans}), 
the inverse square root generates the right combinatorial
factors under differentiation, and the nontrivial $\chi$ dependence
accounts for the commutation relations between $q$ and $p$.
The first few harmonic oscillator energy cumulants may now be 
straightforwardly found via Eq.~(\ref{cumdef}),
\begin{eqnarray}
\langle\langle H_s^2\rangle\rangle 
&=&(1/2)[-(\varepsilon^2/2) + 4 E^2 
- 2 \varepsilon^2 A]\; , \label{2cum}
\\
\langle\langle H_s^3\rangle\rangle  
&=&-(E/2)[-16 E^2 + \varepsilon^2(1+12 A)]\; ,
\\
\langle\langle H_s^4 \rangle\rangle  
&=&48 E^4 - 4  \varepsilon^2 E^2 (1+12 A) \nonumber \\
&+& \varepsilon^4 [(1/8) + 2 A + 6 A^2] \, .
\label{cums}
\end{eqnarray}
After inserting the mean square values for an ohmic bath
(see the discussion above eqs.~(\ref{x},\ref{y})), 
Eq.~(\ref{2cum}) is identical to the main result of Ref. \cite{nb}.

Alternatively, we now consider the diagonal matrix
elements $\rho_{nn}$.
An analytical expression for the density matrix in the energy basis
may be found by using the wavefunctions of the harmonic oscillator,
$\psi_n(q) \propto  e^{-\gamma^2 q^2/2} H_n(\gamma q)$
where $\gamma=\sqrt{m \omega/\hbar}$ and $H_n(x)$ is the $n^{th}$ Hermite
polynomial.
In the energy basis, the density matrix is given by
$\rho_{nm} = \int dq dq' \psi^{\ast}_n(q) \langle q\vert \rho \vert 
q' \rangle  \psi_m(q')$.
 The position space integrals may be done using two
different copies of the generating function for the Hermite
 polynomials.
The diagonal elements may be found by equating equal powers of the
generating variables.  We first define the dimensionless variables
$x = 2 \gamma^2 \langle q^2 \rangle$, 
$y = 2\langle p^2 \rangle/( \gamma^2 \hbar^2)$, and
$D= 1+x+y+x y$.
$x$ and $y$ are related to the major and minor axes of an uncertainty ellipse.
The isolated harmonic oscillator (in it's ground state) obeys two
important properties:  minimum uncertainty (in position and momentum)
and equipartition of energy between average kinetic and potential energies.
The influence of the environment causes deviations from these ideal behaviors
which may be accounted for by introducing two new parameters, 
$a=(y-x)/D,\; b=(x y -1)/D$ with  $-1 \le a\le 1$ 
and $0 \le b\le 1$. The deviation from
equipartition of energy is measured by $a$, while the deviation from
the ideal uncertainty relation is measured by $b$.  
We find
\be
\rho_{nn} = \frac{n!}{2^n} \sqrt{\frac{4}{D}} \sum_{m=0}^n
 \sum_{p=0}^n \frac{(a^2)^p  (2 b)^m}{m! (p!)^2} \delta_{n, m+2 p}\, .
\label{pnn}
\ee
All terms are positive (because $a$ appears only squared),
less than one, and obey ${\rm Tr} \rho=1$. 
The summation over $m$ and $p$ define a polynomial $P_n$ of order $n$ in
$a$ and $b$.  The probability for the lone oscillator
to be measured in an excited state clearly decays rapidly with
level number.
These probabilities also reveal environmental information.
For example, $P_1 =2 b$ and is thus
only sensitive to the area of the state, while
$P_2 = a^2+ 2 b^2$ depends on both the uncertainty and energy
asymmetry. 
Additionally, if we expand the first density matrix
eigenvalue \cite{ms,weiss} with respect to small deviations of
$x$ and $y$, we recover $\rho_{11}$ in agreement with 
our general argument.  
%

Although $x$ and $y$ have been treated as independent variables, the
kind of environment the system is coupled to replaces these
variables with two functions of the 
coupling constant.  For example, with the ohmic bath \cite{weiss,nb}
(in the under-damped limit), the variables are 
\begin{eqnarray}
&& x(\alpha) = \frac{1}{\sqrt{1-\alpha^2}} 
\left(1-\frac{2}{\pi}{\rm arctan}
\frac{\alpha}{\sqrt{1-\alpha^2}}\right)\; ,
\label{x} \\
&& y(\alpha) = (1- 2 \alpha^2) x(\alpha) + \frac{4 \alpha}{\pi} 
\ln\frac{\omega_c}{\omega}\; ,
\label{y}
\end{eqnarray}
where $\alpha$ is the coupling to the environment 
in units of the oscillator frequency and $\omega_c$ is a high frequency cutoff.
This bath information is shown in Fig.~(\ref{bathexcite}) 
with $\omega_c = 10 \omega$.  
The  trajectory of the 
line over the surface shows how the probabilities evolve as the coupling
$\alpha$ is increased from 0 to 1.  Other kinds of environments would trace
out different contours on the probability surface.
 
In conclusion, we have shown that projective measurements
of the system Hamiltonian at zero temperature reveals entanglement
properties of the many-body quantum mechanical ground state.
Consequently, repeated experiments on simple quantum systems
give information about the nature of the environment, the
strength of the coupling and entanglement.  The larger the
energy fluctuations, the greater the entanglement. 
There are several possibilities
for experimental implementations.  We have mentioned
measurement of persistent current \cite{pc,pc2}
as well as projecting on the 
system's energy eigenstates.  Another measurement possibility is
a zero temperature activation-like process \cite{act}
where the dominant mechanism
is not tunneling, but the same quantum effects of the environment
which we have discussed here.

This work was supported by the Swiss National Science Foundation.


\begin{thebibliography}{02}
\bibitem{note2}
$\omega_c$ is the high frequency cutoff needed to regularize
the theory.
\bibitem{note3}
$\rho$ cannot always be written as an effective
thermal distribution. Ref. \cite{ms} explicitly
demonstrates this if the system contains two
or more oscillators.
\end{thebibliography}
\end{document}